# Equilibrium Blocking Model of Isometric Tension

Henry G. Zot[*], Javier E. Hasbun[†], and Nguyen Van Minh[‡]

Departments of [*]Biology, [†] Physics, and [‡]Mathematics
University of West Georgia, Carrollton, GA 30118
hzot@westga.edu

ABSTRACT

Calcium activation of striated muscle is known to exhibit a strongly cooperative dependency on calcium. Because the calcium receptor protein, troponin (Tn) is known to bind calcium non-cooperatively and has yet to be linked to a cooperative change in the myosin-blocking protein, tropomyosin (Tm), we describe a model in which cooperativity is exclusively a myosin-dependent mechanism. The model couples the energies of three well-described reactions with actin, namely, actin-Tn, actin-Tm, and actin-Tm-myosin, to the well-documented positions of Tm, B (blocking), C (central), and M (myosin-dependent) respectively. Results of simulations with and without data are consistent with a strand of Tm composed of ~20 subunits being moved by the concerted action of 3-5 myosin heads resulting in an all-or-none activation of the entire region of the thin filament overlapped by myosin. Equations derived from the model fit both equilibrium myosin binding data and steady-state calcium-dependent tension data and simulate non-cooperative calcium binding both in the presence and absence of myosin. All parameters of the model can be determined experimentally. The mechanism is consistent with steric blocking being both necessary and sufficient for regulation of striated muscle and can be applied to any actin-based contractile system that includes Tm and filamentous myosin.

INTRODUCTION

For vertebrate striated muscle, modeling steady-state isometric tension data with the known properties of calcium binding has proven difficult to achieve. The tension response to varying calcium is distinctly sigmoidal, suggesting an underlying cooperative mechanism. A potential basis for cooperative activation is the association of myosin with thin filaments [1,2]. All present models of striated muscle regulation were derived originally from fitting myosin binding to thin filaments at fixed calcium [2-9]. Given a myosin binding site on each actin monomer, an allosteric mechanism based on Tm controlling seven myosin binding sites has been proposed [1,10]. However, a strictly allosteric model must be reconciled with the muscle lattice, which allows only 1-2 myosin bound per Tm subunit [11,12]. In addition, calcium rather than myosin varies in the muscle. Given these restrictions, cooperative calcium binding has been proposed as a mechanism for activating muscle contraction [8,9]. However, direct measurements of calcium binding have been consistently documented to be non-cooperative both in the presence and the absence of myosin [13,14].

Thin filaments consist of continuous parallel strands of polymeric actin and tropomyosin molecules (c.f. [15] for review). Two strands of the actin polymer bind side-by-side along their length to form a single double-stranded helical structure, and one strand of tropomyosin is located along each side of the actin helix. In striated muscle, one calcium-binding troponin molecule is bound to each molecule of tropomyosin. A linear group of seven actin monomers, a tropomyosin molecule, and a troponin molecule, corresponding in length to the pitch of the actin helix constitute a structural repeat. There are 26 structural repeats, each defined by the length of a Tm subunit, in series along the length of a thin filament (1000 nm). Approximately 20 Tm subunits (75% of thin filament) are overlapped by thick filaments in the muscle lattice at rest.

Thin filament activation, which gives rise to isometric tension, is defined as the exposure of binding sites for interaction with myosin [16]. Owing to an extended structure, each subunit of Tm regulates the interaction of up to seven potential myosin molecules with binding sites of actin (one site per actin monomer). The flexibility of the Tm polymer allows each subunit to occupy any of three discreet positions relative to the location of the myosin binding site on the outer face of the actin helix. Positions B, C, and M of Tm correspond to blocking, central, and myosin-dependent respectively [17]. Tm in Position B completely blocks myosin binding and is favored at low calcium [18]. Increasing calcium shifts the distribution of Tm away from Positions B to favoring Position C [19]. Structural reconstructions of Tm in Position C reveal a partial overlap with the myosin binding site [20], but Tm is expected to undergo extensive thermal motions about Position C, which would expose the myosin binding site [21]. Myosin binding displaces Tm to Position M, regardless of calcium or the starting position of Tm [22].

Each of the positions of Tm can be related to a specific biochemical interaction with actin, namely, actin-Tn, actin-Tm, and actin-Tm-myosin for Positions B, C, and M respectively. For Position B, an interaction between Tn and actin [23] can only occur when the complex of Tm and Tn (Tmn) is located in this position [24]. Calcium binding to Tn weakens the Tn-actin interaction [23], which is consistent with a shift of the distribution of Tm to Position C [19,20] and the displacement of Tn to the inner domain of actin [25]. For Position C, a direct interaction between Tm and actin is consistent with the position of Tm in Tm-decorated actin filaments (no Tn and myosin) and with a favorable orientation of Tm that promotes multiple electrostatic interactions between actin and Tm in this position [20]. For Positions M, the association of myosin enhances the affinity of Tm for actin [1] and Tm enhances the affinity of myosin for actin [2]. These observations are consistent with the formation of a ternary complex of myosin, Tm, and actin. That this complex may induce structural changes in Tm is suggested by reconstructions of myosin decorated thin filaments showing a contiguous length of unsupported Tm in Position M beyond the last bound myosin [22]. A stiffening of Tm in Position M would enable free access to myosin binding sites as has been modeled previously [3]. In summary, a Tm-actin interaction could allow thermal motions of Tm to extend Tmn from equilibrium Position C to non-equilibrium flanking positions. For transient Tm to acquire stability, favorable binding reactions must take place in Positions B and M.

The Hill model [26] stands alone in being capable of explaining a sigmoidal calcium activation. Hill proposed that calcium binding to Tn and myosin binding to actin each contributes to the work associated with the position of Tm [3,26]. Rather than strictly an allosteric mechanism, interactions between Tm subunits were proposed as a second mechanism of enhancing myosin binding [3]. Both calcium and myosin binding perturb nearest neighbor Tm interactions and thereby contribute to the movement of Tm [3,26]. To achieve cooperative activation by a nearest neighbor mechanism, calcium binding must not only promote the movement of Tm but also alter interactions between nearest neighbor Tm subunits [26]. However, experimental evidence for altered interactions between Tm subunits remains to be established [27].

As a result, tension data have not been analyzed with the aid of a comprehensive quantitative model even when the results are purported to be consistent with a nearest neighbor mechanism [28]. To address the possibility that Tn does not regulate interactions between nearest neighbor Tm subunits, we examined whether calcium activation could be based solely on well-established protein interactions of the thin filament. The model would have to provide a biochemical basis for the known positions of Tm and be consistent with the following data: only 1-2 myosin are bound per Tm subunit at maximum calcium, calcium binding is non-cooperative at fixed myosin, the binding of myosin is cooperative at fixed calcium, and activation by calcium is highly cooperative. We demonstrate a model that meets these requirements and fits a challenging set of isometric tension records.

MODEL DESCRIPTION

To address the problem of fitting calcium dependent steady-state isometric data, we develop a model based on the biochemical states required to couple binding energy to the work associated with specific positions of Tm. Because the stability of the Tmn complex in Position B depends on the affinity of the Tn-actin interaction, a discreet coupled state exists for each of the calcium bound state of Tn. Given two regulatory sites for calcium, three actin-bound states of Tn can couple Tmn to Position B, namely, $ATmn.Ca_0(B)$, $ATmn.Ca_1(B)$, or $ATmn.Ca_2(B)$, for Tn with zero, one or two bound calcium respectively. Position C requires only the interaction between actin and Tm, $ATmn.X(C)$ independent of the calcium bound state of Tn (X represents 0, 1, or 2 calcium bound to Tn). The interaction of Tm with Tn does appear to be necessary for Tm to adopt an optimal structure for interaction with actin [29]. Similar to Tm in Position C, stability of Tm in Position M is independent of calcium binding to Tn, but does require energy associated with an interaction, which, in this case, involves a ternary complex among myosin, Tm, and actin represented by the state, $ATmn.X(M)$. Therefore, the system for the regulation of vertebrate striated muscle is composed of five coupled states, $ATmn.Ca_0(B)$, $ATmn.Ca_1(B)$, $ATmn.Ca_2(B)$, $ATmn.X(C)$, and $ATmn.X(M)$.

If calcium binding serves only to weaken the Tn-actin interaction, calcium cannot regulate nearest neighbor interactions with Tm [3,26]. Instead, we inferred a novel cooperative mechanism from the fact that the actin-myosin interaction is the only interaction that has been shown to be cooperative [1,2,4] and from the derivation of the myosin binding relationship (see Eq. 6 below). The novel mechanism we propose is based on the formation of a rigid conformation of Tm and the propagation of this structure along the Tm polymer. The propagation results from two seemingly contradictory properties of Tm, namely, that Tm adopts a rigid structure required for Position M only by coupling simultaneously with multiple myosin and that the rigid structure allows only one myosin at a time to be coupled, rendering all other myosin of a segment unable to couple Fig. 1). Because a single segment can have only one coupled myosin, the multiple myosin required for a super segment must be in separate component segments (Fig. 1). We show that the size of the super segment propagates exponentially with the formation of the coupled state, ATmn.X(M), if the minimum segment size of one Tm subunit is allowed to vary with the probability that ATmn.X(M) will form, designated as $P_M$. The super segment concept is based on the possibility that an axial stretch between periodic points of the linear sequence of mechanically linked Tm subunits drives the conformational change in Tm; a successful stretch requires tension derived from multiple myosin coupling events spread over multiple segments.

$P_M$ depends on the stability of the segments. As coupling decays over time and because the segment excludes simultaneous coupling by more than one myosin, the stability of a segment depends on myosin being available within the segment to accept coupling dynamically. Given a stochastic association of myosin, the myosin available for coupling depends critically on the average number of myosin bound, which is 1.3 myosin per Tm subunit for insect flight muscle [11]. In stabilizing the coupled state, the pool of available myosin has the properties of free ligand, which we refer to as *free myosin*. Although any of the myosin bound within a segment could be free, the pool of free myosin is limited to one myosin per Tm subunit by the rigid structure of Tm and the coupling reaction. An expanding segment increases the stability of coupling by increasing the free myosin (Fig. 1). Maximum expansion of a segment is determined by the ability of unsupported Tm in Position M to resist the tendency to return to equilibrium in Position C, and, hence is an intrinsic property of Tm. Similarly, the intrinsic property of Tm determines the number of coupled myosin in separate segments required to induce the rigid conformation of the super segment. Thus, free myosin is required for $P_M$ to approach one. The blocking of myosin by Tm in Position B regulates $P_M$ in vertebrate striated muscle. Where Tn is absent (smooth and non-muscle), $P_M$ could be regulated by controlling the affinity of myosin for actin.

Given that the myosin binding site of actin is fully blocked by Tm in Position B and fully exposed when Tm is in Position M [17], activation of the thin filament is functionally *off* when Tm is in Position B and functionally *on* when Tm is in Position M. However, the question arises about the status of thin filament when Tm is in Position C. In our model, isolated units of the thin filament in which Tm is in Position C should have biochemical properties similar to the actin-Tm complex. As the kinetics of myosin binding to either Tm-actin or pure actin are indistinguishable [16], we propose that the thin filament in Position C supports the cycling of myosin for sliding filaments as would be found in isotonic conditions. The displacement between thin and thick filaments by sliding filaments is expected to destabilize the interaction between myosin and actin and, hence, Position M. Thus, isometric conditions of the muscle are required to maximally enhance the affinity of coupled myosin by slowing the catalytic rate, thereby stabilizing the formation of the super segment and maximizing free myosin.

MODEL DERIVATION

Conservation Equations

To derive the conservation of mass relationships, let $U_T$ represent the total number of Tm subunits in a given preparation. If the number of Tm subunits (U) per segment (S) is defined with a parameter $g$, ($g \equiv U/S$), then the accounting of Tm subunits distributed among the states of the system is given by

$$U_T = ATmn.Ca_0(B) + ATmn.Ca_1(B) + ATmn.Ca_2(B) + ATmn.X(C) + g(ATmn.X(M)) \qquad (1)$$

where $S_T = U_T/g$ represents the total number of segments in a preparation. Letting $ATmn.Ca_0(B)/U_T$, $ATmn.Ca_1(B)/U_T$, $ATmn.Ca_2(B)/U_T$, $ATmn.X(C)/U_T$, and $ATmn.X(M)/S_T$ be $B_1$, $B_2$, $B_3$, $C$, $M$ respectively, the modeling results can be applied to any preparation and Eq. 1 becomes

$$1 = B_1 + B_2 + B_3 + C + M \qquad (2)$$

From one coupled myosin per segment (Fig. 1), it can be seen that $ATmn.X(M) = S$ and $S/S_T = M$.

The reactions that give rise to Eq. 2 (Fig. 2) also involve states of Tn and myosin that do not contribute to the positioning of Tm. To account for the uncoupled mass of Tn, we let $T_1 = Tn.Ca_0/U_T$, $T_2 = Tn.Ca_1/U_T$, and $T_3 = Tn.Ca_2/U_T$ represent the calcium bound states of Tn (0, 1, and 2 calcium bound respectively) tethered by Tm in Positions C and M. Substituting for C and M in Eq. 2 gives the relationship for the conservation of Tn,

$$1 = B_1 + B_2 + B_3 + T_1 + T_2 + T_3 \qquad (3)$$

Not all of the mass of muscle myosin can be coupled to the position of Tm at any given time owing to catalytic intermediates of myosin that cannot bind strongly to actin and to unfavorable orientations of thick filament bound myosin. Letting $Y_T$ represent the total mass of myosin in the muscle and $M_T \leq Y_T$, then $M_T$ represents the mass of myosin capable of being coupled to the work associated with Tm in Position M. If $m_f$ represents the free myosin capable of being coupled, then the following relationship is given

$$M_T = m_f + ATmn.X(M) \qquad (4)$$

Eq. 4 accounts for all non-extraneous biochemical and structural intermediates of myosin. The expression, $M_T/S_T = m_f/S_T + ATmn \cdot X(M)/S_T$ accounts for the fraction of $M_T$ present in segments that have formed. Given one coupled myosin per Tm subunit (Fig. 1), $M_T$ is a constant, given by $M_T = U_T$; hence, $M_T/S_T = g$ and, thus substituting for the expression above gives

$$g = m + M \tag{5}$$

where $m = m_f/S_T$, represents the fraction of free myosin that formed by segments.

Equilibrium Relationships For Position M
In addition to m, the formation of the super segment depends on the fraction of C that is transiently available (inset Fig. 2). The latter is given by $C/K_A$, where $K_A$ represents the energy required to displace Tm from its equilibrium position (Position C) for interaction with myosin in Position M (Fig. 2). As $C/K_A$ and m are independent, the reaction proceeds by mass action given by $S_S = K_0 C m^n$ where $K_0 = (K'_0 U_T)^n / K_A$, and $K'_0$ and $K_A$ are intrinsic constants, $S_S$ represents the fraction of Tm in Position M that formed by super segments, and the parameter, $n$, represents the number of myosin that coupled simultaneously by super segments. Letting SS and $SS_T$ represent a super segment and the total mass of super segments respectively, then $S_S = SS/SS_T$. As $S = nSS$, there is a proportional relationship between segments and super segments, hence $S/S_T = SS/SS_T$. Because $M = S/S_T$ (see above), $S_S = M$ (Fig. 2), from which the following is derived

$$M = K_0 C m^n \tag{6}$$

From Eq. 5, we substitute for m in Eq. 6 to get

$$M = K_0 C (g - M)^n \tag{7}$$

To account for variable segment size, we make g dependent on $P_M$ and a parameter $\alpha$ that determines the maximum segment length in subunits of Tm, i.e., $g \equiv 1 + \alpha P_M$. When $P_M = 1$, g is a constant, $g_{max}$, where $g_{max} = 1 + \alpha$, and the maximum super segment length is given by $n(g_{max})$.

Because M represents the fraction of Tm subunits that are in Position M, M is also a Bayesian probability that myosin will be coupled to the position of Tm. Thus, $P_M = M$, which by substitution above gives

$$g = 1 + \alpha M \tag{8}$$

Substituting Eq. 8 into Eq. 7 results in a relationship that can be evaluated given experimental parameters $n$ and $\alpha$,

$$M = K_0 C (1 + \alpha M - M)^n \tag{9}$$

Equilibrium Relationships For Position B

The expressions for calculating the fraction of regulatory units with Tm in the blocking position are derived from a two-step sequence that involves both a calcium-independent transition governed by $K_B$ and an interaction between Tn and actin governed by one of three possible stability constants, $K'_1$, $K'_3$, and $K'_5$ (Fig. 2). As Tn cannot bind actin in Position C, the transition between B and C is equivalent to an independent process that controls the mole fraction of actin binding sites for Tn. Thus, the fraction of actin available for interaction with Tn in Position B is independent of calcium and Tn, and is given by $C/K_B$, which is the fraction of Tm transiently present in Position B (inset and outset, Fig. 2). The formation of a given B complex ($B_1$, $B_2$, or $B_3$) requires both $C/K_B$ and the corresponding tethered state of Tn, $T_1$, $T_2$, of $T_3$, respectively. These equilibria can be represented by

$$B_1 = K_1 C T_1 \tag{10}$$

$$B_2 = K_3 C T_2 \tag{11}$$

$$B_3 = K_5 C T_3 \tag{12}$$

where $K_1$, $K_3$, and $K_5$ are all first order, i.e., $K_1 = K'_1 U_T / K_B$, $K_3 = K'_3 U_T / K_B$, and $K_5 = K'_5 U_T / K_B$.

Similarly, the expressions for calculating the fraction of regulatory units with calcium associated with Tn are derived from Tn in actin-bound and tethered states (Fig. 2):

$$T_2 = K_2 T_1 \tag{13}$$

$$T_3 = K_2 T_2 \tag{14}$$

$$B_2 = K_4 B_1 \tag{15}$$

and

$$B_3 = K_4 B_2 \tag{16}$$

where $K_2$ and $K_4$ are defined as $K'_2 Ca$ and $K'_2 Ca$, respectively. Calcium, Ca, input is the only independent variable (Table 1).

Fitting Mutant Troponin Data
We test the ability of our model to simulate the results of replacing wild-type Tn with mutant Tn unable to bind calcium [28]. To model the replacement of wild-type Tn, we introduce $ATmn^-(B)$ and $Tn^-$ to represent states that contain mutant Tn. Letting $B^-$ and $T^-$ represent $ATmn^-(B)/U_T$ and $Tn^-/U_T$, respectively, in the following,

$$1 - p = B^- + T^- \tag{17}$$

where the parameter, *p*, represents the fraction ($0 \leq p \leq 1$) of total Tn that is wild type. For the derived mass action relationship, $B^- = K_1 C T^-$, we use the same actin binding constant for mutant and wild-type apo-Tn ($K_1$), although this is a simplifying assumption without experimental support. By substitution into Eq. 17, we obtain the following relationship for evaluation,

$$T^- = 1 - p - K_1 C T^- \tag{18}$$

The conservation equations for coupled states and wild-type Tn states are given respectively by,

$$1 = C + M + B_1 + B_2 + B_3 + B^- \tag{19}$$

$$p = B_1 + B_2 + B_3 + T_1 + T_2 + T_3 \tag{20}$$

Calculation of Calcium Activation

To minimize the number of simultaneous equations to solve, we substituted equivalent expressions (Table 1) into Eqs. 19, and 20, to give

$$C = (1 - M) / (1 + (1 + K_4(1 + K_4))K_1 T_1 + K_1 T^-) \tag{21}$$

and

$$T_1 = p - ((1 + K_4(1 + K_4))K_1 C + K_2(1 + K_2))T_1 \tag{22}$$

The solution of Eqs. 9, 18, 21 and 22 for an arbitrary calcium concentration yields values for variables C, M, $T_1$, and $T^-$. From these values, all other variables are evaluated using the relationships in Table 1.

Expressions to Fit Myosin Binding Data

We develop the relationships necessary to fit the cooperative binding of myosin as detected by a change in fluorescence at low fixed calcium [16]. The myosin-dependent fluorescence change is represented by two sequential reactions, $B_1 \rightleftharpoons C + T_1$ and $C + m_f \rightleftharpoons M$, where $m_f$ is the measured free myosin concentration and $K_1$ and $\gamma K_0$ govern the first and second reaction respectively; the factor, $\gamma$, relates $m_f$, which has maximum degrees of freedom in solution, to myosin constrained by the muscle lattice. The fluorescence change is assumed to result from an increase of $T_1$ by the first reaction, in response to myosin association in the second reaction, which is consistent with a direct correlation between the fractional fluorescence change, $\Delta F$, and M. Letting $\Delta F = M$ and $M = 1 - B_1$, the relationship between the fluorescence change and free myosin is given by

$$\Delta F = \gamma K_0 / K_1 (1 + (\alpha - 1)\Delta F)^n [m_f]^n ((1 - \Delta F)/\Delta F) \tag{23}$$

To fit the fluorescence data, this equation presents the formidable problem of solving an $n^{th}$ order polynomial. This is avoided by reversing the dependent variable of the data and transforming Eq. 23 accordingly

$$[m_f] = (1 + (\alpha - 1)\Delta F)^{-1}(\Delta F^2/(\gamma K_0/K_1(1 - \Delta F)))^{1/n} \qquad (24)$$

Equation 24 is used to derive $\gamma K_0/K_1$ from myosin-dependent fluorescence data, given values for $n$ and $\alpha$.

Total Myosin Binding

In our model, total equilibrium binding of myosin ($\Theta$) is the sum of cooperative ($\Delta F$) and non-cooperative ($\theta$) binding, given by

$$\Theta = \Delta F(1 + (7(\alpha - 1) \Delta F + 6)\theta)/7(\alpha + 1) \qquad (25)$$

For this equation, $\Delta F$ is generated by Eq. 23 and $\theta$ is generated by the following binding isotherm,

$$\theta = K_f[m_f]/(1 + K_f[m_f]) \qquad (26)$$

where $K_f$ is the measured association constant for strong binding myosin with pure actin.

RESULTS

A number of observations place constraints on the parameters of the system ($K_0$, $K_1$, $K_3$, $K_5$, $K'_2$, $K'_4$, $\alpha$, and $n$). To establish values for $K'_2$ and $K'_4$ from published transient calcium binding measurements [14], we paired the fastest measured on-rate with the two measured off rates. Thus, based on the ratio of measured rates (association/dissociation), $K'_2 = (2.5 \times 10^7 \text{ M}^{-1}\text{s}^{-1}/15 \text{ s}^{-1}) = 1.67 \times 10^6 \text{ M}^{-1}$ and $K'_4 = (2.5 \times 10^7 \text{ M}^{-1}\text{s}^{-1}/150 \text{ s}^{-1}) = 1.67 \times 10^5 \text{ M}^{-1}$ (Table 2). From the ratio, $K'_2/K'_4$ and conservation at equilibrium (Fig. 2), values for $K_1/K_3$ and $K_1/K_5$ can be established (Table 2). This leaves $K_1$ as the only adjustable parameter in the absence of myosin (Fig. 2).

The myosin dependent parameters can be estimated from structural considerations. Certain favorable reconstructions of myosin-decorated thin filaments reveal 120-300 nm of unsupported Tm in Position M [22], which corresponds to 4-7 Tm subunits (based on 38.7 nm per Tm subunit). We take this range as the possible values of $g_{max}$ (Table 2), from which we establish $\alpha$ to be $3 \leq \alpha \leq 6$. The range of $n$ is constrained by the fact that $n$ segments of length $g_{max}$ cannot exceed the number of Tm units within the overlap region of a single thin filament, which is ~20. Hence $n \leq 5$ (Table 2).

We studied the behavior of the model restricted to Positions B and C, where the fraction of Tm in Position C is taken as equivalent to activation in the absence of myosin. In these simulations we solved for C using Eqs. 9, 18, 21 and 22 given an arbitrary calcium.

To prevent Position M, we set $K_0$ to zero, which means $K_1$ became the sole adjustable parameter of the simulation. Three characteristics of the curves vary with $K_1$, namely, the baseline (activation at lowest calcium), the extent (difference between baseline and plateau), and sensitivity to calcium (calcium concentration at the midpoint between minimum and maximum). As $K_1$ is decreased, the fractional activation becomes less sensitive to calcium and the baseline decreases (Fig. 3, inset). The extent of activation reaches a maximum near Curve C (inset, Fig. 3), which is the simulation consistent with the results of particle tracking [19]. Biochemical results estimate the range of activation to be approximately 5% (baseline) – 50% (plateau) [30], which is given by Curve E (inset, Fig. 3). In summary, the simulations without myosin showed that calcium activation is not cooperative, calcium sensitivity is not uniquely dependent on calcium binding, the full range of calcium activation cannot be achieved in the absence of myosin given physiological constants for calcium binding, and the extent of calcium activation may be < 50% to achieve a low baseline. For subsequent simulations, we used the conditions that generated a low baseline (Curve E) to investigate the role of myosin in expanding the extent of activation.

We asked whether a cooperative response to calcium could be achieved if myosin were include in the calculation. We simulated this with our model by solving for the sum of C and M using Eqs. Eqs. 9, 18, 21 and 22 given an arbitrary calcium. To include myosin in the simulation, we set $K_0$ to 1 (Table 2). We tested the two parameters, $\alpha$ and $n$ (Eq. 9) individually for cooperative activation. Of the two parameters, $\alpha$ is seen as crucial, because, compared with a non-cooperative response (Curve 1; Fig. 3), a cooperative response is not observed for $\alpha = 0$ even when $n = 6$ (Curve 2; Fig. 3), whereas a cooperative response is achieved with $\alpha = 20$ and $n = 1$ (Curve 3; Fig. 3). A synergistic increase in steepness of the curves results from the combination of $n > 1$ and $\alpha > 2$ (Curves 4–8; Fig. 3). For a given $K_0/K_1$ and $\alpha$, increasing $n$ generates a progressively steeper calcium activation, greater extent of calcium activation, and greater calcium sensitivity (Curves 5, 7, 8; Fig. 3). By increasing $K_0/K_1$, given constant $n$ and $\alpha$, the curves shifts toward greater calcium sensitivity and extent of activation, but the steepness does not change significantly (Curves 4–6; Fig. 3). Tm in Position C declines to zero as calcium is increased (Curve 0; Fig. 3). Thus, in simulations of isometric contraction, Position C contributes only to baseline activation and most of the transitions are between Positions B and M.

We investigated combinations of $n$ and $\alpha$ that generates the greatest extent of activation. Approximately 90% of the full extent of activation (~5% to ~95%) is achieved not only with $n = 4$ and $\alpha = 4$ (Curve 7; Fig. 3), but also with $n = 3$ and $\alpha = 5$ and $n = 5$ and $\alpha = 3$ (not shown). The maximum super segment ($n(g_{max})$; Table 2) computed from these pairs of parameters converges on a value of ~20 Tm subunits. Further investigation showed that the curves become too steep to calculate using standard software packages when $n(g_{max}) > 21$, and the extent of activation becomes progressively reduced when $n(g_{max}) < 19$. These results compare favorably with the region of thin filament overlap with thick filaments estimated to be ~20 Tm subunits.

Mutant Tn incapable of binding calcium is shown not only to reduce maximum tension, as expected, but to also reduce the steepness and calcium sensitivity of the curves generated by steady-state additions of calcium [28]. Given the simulations above, we asked whether all the effects of mutant Tn could be reproduced with our model. Fitting data from native muscle fibers required combinations of $n$ and $α$ in which $n(g_{max})$ is 20 as noted above, owing to the steepness of the steady-state tension response to calcium (diamond, Fig. 4). To simulate mutant Tn that cannot bind calcium, we introduced a new coupled state of Tn, $B^-$, which binds actin identically with native Tn devoid of calcium ($K_1$) and an uncoupled state of Tn, $T^-$, which is in equilibrium with $B^-$ (Table 1). Given the fraction of wild-type Tn ($p$; Table 2) as the only adjustable parameter, the model accurately simulated the calcium sensitivity and steepness of the data (Fig. 4). Although simulations were consistent with a reduction in maximum tension, the model overestimated the amount of wild-type Tn required to achieve maximum tension, by 5, 10, and 13% for reconstitutions with 80, 60, and 20% wild-type Tn, respectively (Fig. 4). However, by reducing the actin binding constant of mutant Tn ~50%, all but the 20% data were fit to the accuracy of the measurement without overestimation (not shown). The affinity of mutant Tn for actin was not measured [28].

To further test our model, we simulated myosin binding data [16]. Our working assumption was that the published fluorescence data [16] represents myosin coupling, because the movement of Tm to Position M accompanies the fluorescence change. Using Eq. 24, we obtained a best visual fit of the fluorescence data (circles, Fig. 5) with $3 \times 10^{-3}$ as the value of the dependent variable $γK_0/K_1$. The constant, $γ$, accounts for myosin being freely diffusible rather than constrained by the muscle lattice. Nevertheless, in the above simulations of the intact muscle, $K_0/K_1$ is typically $2 \times 10^{-3}$ (1/500) demonstrating only a small discrepancy in the two results. Total myosin binding in our model is the sum of two distinct populations of myosin, namely, coupled and non-coupled. Coupled myosin binds cooperatively and is given by the change in fluorescence ($ΔF$; Eq. 23). The binding of coupled myosin exposes the bulk of the actin binding sites for association with non-coupled myosin, which binds non-cooperatively ($θ$; Eq. 26). To simulate the behavior of the two populations of myosin, a total binding isotherm ($Θ$; Fig. 5) is generated with Eq. 25, given inputs by both $ΔF$ and $θ$. Best fits of the experimental data were achieved with $n > 3$ owing to the steepness of the response (data not shown), although data confidence is not sufficient to exclude any fits we obtained for $3 ≤ n ≤ 5$.

Calcium binding to regulated actin has been shown to be non-cooperative both in the absence and presence of myosin [13,14]. Calcium binding in our model is the sum of all calcium bound states of Tn, i.e., $B_2$, $B_3$, $T_2$, and $T_3$ (Table 1), for arbitrary free calcium. The total mass of Tn is represented at equilibrium in the absence of myosin by all four calcium bound states plus $B_1$ and $T_1$, but the myosin-dependent movement of Tm to Position M shifts all of the Tn mass to $T_1$, $T_2$, and $T_3$ (Fig. 2). Although the shift in Tn mass results in a change in calcium affinity from a mixture of $K_2$ and $K_4$ to exclusively $K_2$, no cooperative binding is expected. This is confirmed by simulations of calcium binding at zero and saturating myosin (inset, Fig. 5).

DISCUSSION

A set of equilibrium-based equations are shown here to fit disparate data related to muscle regulation. The equations are derived from a single thermodynamic system of five states for fast twitch muscle; one fewer states would be required to specify cardiac and slow twitch muscles, which have only one calcium regulatory site. The purpose of the system is to couple biochemical binding energy to the movement of Tm among three discrete positions. Although the system is open, inasmuch as mass is exchanged between coupled and uncoupled Tn and coupled and free myosin, the total mass of the system and the surroundings is not only fixed but also spatially constrained by the muscle lattice. We show that the fractions (or probabilities) of all states can be computed simply from one set of simultaneous equations.

Free myosin was previously considered constant (c.f., Fig 5, [26]), owing to the constraint of the muscle lattice, but free myosin propagates in the model we describe. By inspection of Eq. 6, free myosin can be seen as having a possible exponential relationship with myosin and Tm in Position M, only if free myosin is variable. This fact and the possibility that free myosin can be bound to actin and unregulated with regard to generating force for contraction, provide clear advantages to the mechanism we describe. The super segment mechanism is consistent with all present day actin-based contractile systems with or without Tn, including striated muscle, smooth muscle, and non-muscle, because all include Tm and filamentous myosin (myosin II). It is interesting to note that the calcium dependent steady-state tension relationship is sigmoidal for the molluscan adductor muscle [33], even though this muscle does not contain Tn and is regulated by calcium binding directly to myosin. Provided that calcium binding to myosin regulates myosin binding to actin, our model is consistent with $P_M$ being determined by the fraction of molluscan myosin bound by calcium. A similar mechanism holds for more typical contractile systems regulated by the phosphorylation of myosin.

It is difficult to imagine how the cooperative mechanism we propose could have evolved without thick filaments to deliver myosin to the locations of each Tm subunit simultaneously. Given an ancestral Tm gene, Tn could have evolved after the cooperative mechanism strictly as a means of stabilizing the blocking position of Tm. Our model is consistent with the steric blocking theory, which proposes that the position of Tm regulates muscle contraction by blocking myosin binding [31,32]. In addition, the blocking position of Tm competes directly with Tm required for segment formation, thereby destabilizing the super segment. Because the super segment supports free myosin, steric blocking regulates free myosin in addition to coupled myosin.

The requirement for a super segment follows from experimental observation and our simulations of cooperative calcium activation. Although a sigmoidal dependence on calcium can be achieved for a single segment (Curve 3; Fig. 3), experiments show that a single myosin cannot move the entire Tm strand of a thin filament and, instead, demonstrate that the likely length of single segment is represented by 3-7 Tm subunits [22]. To be consistent with this segment size and with the steepness of experimental activation curves, our simulations required simultaneous coupling by 3-5 myosin.

Because simultaneous coupling is excluded from a given segment, then multiple segments must be required to couple multiple myosin simultaneously to the position of Tm, hence the formation of a super segment. The length of this super segment is determined by the number of coupled myosin multiplied by the number of Tm subunits of a component segment. A super segment equal in length to the overlap region of the thin filament (~20 Tm subunits) was generated by the best fit of experimental data.

Coupled and free myosin likely cycle differently with the thin filament in the steady-state. The cycling rate of coupled myosin should slow in order to form a stable ternary complex with Tm and actin. Given 3-5 Tm subunits coupled to myosin, the remaining 80% of a thin filament strand (based only on the overlap region) must contain 15-17 free myosin, which cycle without constraint. This would represent a lower limit, because multiple myosin are likely to be cycling within the domain of a Tm subunit in order to achieve at least one myosin that can be coupled at all times; evidence from insect flight muscle suggests the average number of attached myosin may be 1.3 per Tm subunit [11]. It is possible that the energy required for myosin to couple to the work associated with the position of Tm would prevent coupled myosin from generating force for isometric tension; nevertheless, no such restriction would apply to the bulk of the myosin cycling within a super segment

There is general agreement that calcium shifts the equilibrium away from the blocking position of Tm [4,6,26], but ours is the only model to suggest that the movement to Position C is passive. Based on myosin binding data, others have modeled the transition away from the blocking position as an unfavorable conformational change that requires stabilization by calcium bound to Tn. A recent reconstruction of negatively stained complexes of actin-Tm reveals that cardiac Tm occupies both Positions B and C when Tn is not associated and that association with Tn stabilizes Tm in either Positions B or C when in the absence or presence of calcium respectively [29]. These findings demonstrate a conformational change associated with the formation of the complex between Tn and Tm and are consistent with the possibility that calcium drives a change in conformation to Position C. We addressed the other possibility, namely, that the conformational change in Tm brought about by association with Tn is simply required to stabilize the equilibrium position of Tm in Position C, which is the position shown for Tm by x-ray diffraction of actin-Tm gels in solution [20]. By our model, Tn is uncoupled from the work associated with the position of Tm by the passive movement of Tmn away from Position B, which is made more likely by the weakening of the Tn-actin interaction when calcium binds Tn.

Our simulations are the first to suggest that a balance of protein-protein interactions can determine the calcium sensitivity of muscle activation. Even without myosin, the calcium sensitivity of the transitions between Positions B and C are seen to depend on the actin-Tn interaction ($K_1$; inset, Fig. 3). This effect is based on the reaction that governs the transition of Tm between Positions B and C, which can be seen by comparing our model of actin-Tn formation by bimolecular reaction with previous models of a conformational change. In the absence of calcium, it can be shown that the fraction of Tm in Position C is a simple proportion, $C = 1/(1 + K_1)$, if a conformational change is assumed, but is a quadratic, $K_1C^2 + C - 1 = 0$, if the reaction is considered bimolecular; similar relationships can be derived for infinite calcium and are presumed to occur for arbitrary calcium. Thus, the reduced calcium sensitivity of muscle fibers regulated by mutant Tn is explained by our model as an inability of mutant Tn to weaken its interaction with actin. We also observed parallel shifts in calcium sensitivity when $K_0/K_1$ was altered (Fig. 3), which simulates the effect that changing physiologic and signal transduction conditions may have on the affinity of myosin for actin.

We suggest the following as potential concerns for the validity of our model. First, the length of Tm that overhangs myosin-decorated actin is the main evidence for a conformational change that stiffens Tm [22]. Further refinements of existing images or additional structural reconstitutions may reveal hidden myosin, as has been suggested [7]. Second, unless present particle-tracking measurements underestimate the fraction of Tm in Position B, our model would have to be modified. We can account for either the particle tracking data (Curve C; inset, Fig 3) [19] or biochemical measurements of activation (Curve E; inset, Fig. 3) [14,16]. To account for both, the characteristics of Position C would have to be changed, but not necessarily the mechanisms describing the transitions between Positions B and C and Positions C and M. It should be noted that Tm in Position C does not play a large role in simulations of isometric conditions (Curve 0, Fig. 3). Third, the rate of Tm movement would have to exceed 1000 $s^{-1}$ to not limit the expected rates of myosin recruitment and transitions between Positions C and M. We are encouraged that present measurements do support rapid transitions of Tm in response to myosin binding [34], but additional study is required to establish upper limits of rate. Additionally, a stiffened conformation of Tm would have to relax fast enough to not limit the decay of Position M. Fourth, a grossly inhomogeneous myosin distribution along the thin filament is the greatest concern for our model of isometric contraction, but variance in bound myosin is also a concern. Reconstructions of insect flight muscle demonstrate a periodicity of bound myosin that corresponds to the central location of each Tm subunit [11]. Statistical variation about 1.3 myosin heads per Tm subunit is low for the planar packing of thin and thick filaments of insect muscle [11] and may even be less for vertebrate muscle, which has trigonal symmetry. Fifth, the model presented here would be inconsistent with activation greater than that achievable by actin and Tm alone. Sixth, for our model to be valid, the affinity and kinetics of calcium binding to Tn in Positions C and M must be the same as those exhibited by pure Tn. To avoid ambiguity in assigning rate constants measured from two calcium binding sites [14], Tn with only one regulatory site would be required. Finally, we suggest that as future experiments become available, it will be possible to infer forward and reverse rates for all of our model's equilibrium constants.

ACKNOWLEDGMENTS

This work was supported by NSF grant MCB-0508203 (HGZ).

FIGURE LEGENDS

Figure 1.
Novel myosin-based cooperative mechanism for vertebrate striated muscle. The diagram depicts non-equilibrium positions of Tm generated by interactions of Tn and myosin with actin. An interaction between Tn and actin (not depicted) is coupled to stabilizing Tm in Position B (▭), which blocks the association of myosin (◊) with actin (•••••). An interaction of myosin with actin and Tm (◣) is coupled to a stiffening of Tm (▬) and stabilization of Tm in Position M. Each coupled myosin is associated with a functional unit of Tm composed of one or more Tm subunits, which is referred to as a segment (S). The stiffening of Tm required to form S requires multiple myosin to be coupled. Since S can have only one coupled myosin, the conformational change in Tm requires the cooperation of multiple S to form a larger functional unit of Tm, referred to as a super segment (SS). The rigid structure of Tm prevents more than one myosin of a segment to be coupled. Only those myosin that have the potential to be coupled within a segment of Tm have the property of a free ligand in a mass action reaction, referred to as free myosin (◇). Free myosin stabilizes the coupled state of myosin by being available to be coupled, as coupling within the segment is dynamic. The number of Tm subunits per S depends on the probability that myosin can be coupled ($P_M$); calcium increases $P_M$ by weakening Position B. The maximum number of Tm per S and the number of myosin that must be coupled to form SS are intrinsic properties of Tm arbitrarily chosen to be 4 and 3 respectively for this diagram.

Figure 2.
Annotated equilibrium model. Three states of Tn coupled to the position of Tm are $B_1$, $B_2$, and $B_3$ (▭,▭,▭) with zero, one, and two calcium bound to Tn (○,▫,▦), respectively. In equilibrium with $B_1$, $B_2$, and $B_3$ are three corresponding states, $T_1$, $T_2$, and $T_3$, which represent Tn tethered to Tm but dissociated from actin in Position B( ▭,▭,▭). Calcium bound to tethered Tn is undefined (▦) for Tm in all but Position B. Position M requires the formation of super segments ($S_s$), which depends on the coupled state of myosin (M). $S_s$ and M require the coupling of $n$ myosin (◣) to the formation of $n$ segments (S). Formation of each segment requires the coupling of myosin binding energy to a non-equilibrium position and conformation of Tm (▬). The stability of a segment depends on the fractional free myosin (m, ◊), which determines the length of S given as $1+\alpha P_M$, where $\alpha$ represents the number of unsupported Tm subunits of a segment and $P_M$ is the probability that a myosin is coupled. Tm transiently present in Position M and Position B, given by $C/K_A$ and $C/K_B$ respectively, are noted as non-equilibrium states (bracket) justified in the inset. Tm in Positions C and M supports cycling myosin intermediates (◊) for sliding filaments and isometric tension respectively. Inset. Schematic representation of Tm energy as a function of position. A direct interaction between Tm and actin gives rise to the only equilibrium position of Tm in Position C. The instability of this position allows thermal motions to carry Tm to Positions B and M ($C/K_A$ and $C/K_B$), where other interactions take place. Position B is stabilized by calcium dependent interactions between Tn and actin and Position M is stabilized by interactions between actin, myosin, and Tm.

Figure 3.
Factors that determine cooperative activation by calcium. Activation is calculated as the fraction of Tm in Positions C and M. The sum of the dependent variables C and M (Table 1) is determined by solving Eqs. 9, 18, 21, and 22 given arbitrary calcium. <u>Inset.</u> Non-cooperative fractional activation in the absence of myosin. Myosin is excluded by setting the parameter $K_0$ (Table 2) to zero, hence the fraction of Tm in Position C is plotted as a function of calcium. <u>Inset adjustable parameters:</u> $K_1$=1 (Curve A); $K_1$=10 (Curve B); $K_1$=20 (Curve C); $K_1$=50 (Curve D); $K_1$=500 (Curve E). <u>Outset.</u> Myosin induces cooperative fractional activation. All curves except Curve 1 include myosin contribution by setting the parameter, $K_0$, to one; for visual comparison to a non-cooperative activation, Curve 1 is a reproduction of Curve E (inset). Curves 2, 3, 5, 7, and 8 illustrate the effects of parameters that control cooperativity: Curves 2 and 3 compare the effects of varying $\alpha$ and $n$ given fixed $K_0/K_1$ and Curves 5, 7, and 8 compare the effects of varying $n$ given fixed $\alpha$ and $K_0/K_1$. For constant $n$ and $\alpha$ (Curves 4-6), increasing $K_0/K_1$ shifts the curves toward greater calcium sensitivity while the steepness remains nearly the same. Curve 0 shows the fraction of Tm in Position C as a function of calcium. <u>Outset adjustable parameters:</u> $K_1$=500, $K_0$=0 (Curve 1); $K_1$=500, $K_0$=1, $n$=6, $\alpha$=0 (Curve 2); $K_1$=500, $K_0$=1, $n$=1, $\alpha$=20 (Curve 3); $K_1$=1000, $K_0$=1, $n$=3, $\alpha$=4 (Curve 4); $K_1$=500, $K_0$=1, $n$=3, $\alpha$=4 (Curve 5); $K_1$=250, $K_0$=1, $n$=3, $\alpha$=4 (Curve 6); $K_1$=500, $K_0$=1, $n$=4, $\alpha$=4 (Curve 7); $K_1$=500, $K_0$=1, $n$=5, $\alpha$=4 (Curve 8). <u>Outset constants:</u> $K'_2$=1.67 × 10$^6$ M$^{-1}$, $K'_4$=1.67 × 10$^{-5}$ M$^{-1}$.

Figure 4.
Fit of isometric tension data. Tension data are taken from [28]. The symbols represent the fractional change in isometric tension of skeletal muscle fibers reconstituted with a mixture of wild-type Tn and mutant Tn unable to bind calcium; the fraction of wild-type Tn is indicated as, ◆ (100%), ■ (80%), ▲ (60%), ● (20%), ✶ (15%). Theoretical curves represent the fraction of Tm in Positions C and M, which is a measure of fractional activation. C and M are determined for arbitrary calcium by solving Eqs. 9, 18, 21, and 22. We normalized the raw simulations by subtracting the baseline (value at lowest calcium) and setting the maximum value (100% wild-type Tn at saturating calcium) equal to 1. The raw simulation with 100% wild-type Tn appears in Fig. 3 (Curve 7). Curves from left to right were generated with the following percentages of wild-type Tn: (left to right) 100% (p=1), 83% (p=0.83), 70% (p=0.7), 33% (p=0.33), 15% (p=0.15). Adjustable parameters used: $p$. Constants used: $K_0$=1, $K_1$=500, $n$=4, $\alpha$=4, $K'_2$=1.67 × 10$^6$ M$^{-1}$, $K'_4$=1.67 × 10$^5$ M$^{-1}$.

Figure 5.
Relationship of IANBD fluorescence data and total myosin binding. All data are replotted from Trybus and Taylor [16]. Fluorescence data (circles) are fit manually by trial and error with Eq. 24, given $\alpha$=3 and $n$=4; the curve through the data is generated by Eq. 23 using $\gamma K_0/K_1$=3 × 10$^{-3}$. Total myosin binding data (squares) are fit with a curve representing the sum of coupled and free myosin binding using Eq. 25. As inputs to Eq. 25, coupled myosin binding is given by the change in fluorescence generated by Eq. 23 ($\gamma K_0/K_1$=3 × 10$^{-3}$) and the free myosin binding is generated by simple mass action ($K_f$=1.67 × $10^7$ $M^{-1}$ [4]; Eq. 26).  <u>Inset.</u> Simulated calcium binding to Tn is non-cooperative. The sum of $B_2$, $B_3$, $T_2$, and $T_3$ (Table 1), which represents the total calcium bound to Tn, is plotted on the Y-axis.  Values for these dependent variables were determined by solving Eqs. 9, 18, 21, 22 (p=1) for arbitrary calcium.  Total calcium binding with zero myosin (–M) and saturating myosin (+M) was simulated using $K_0$=0 and $K_0$=5000, respectively.  Fixed inset parameters: $K_1$=500, $\alpha$=4, $n$=4, $K'_2$=1.67 × $10^6$ $M^{-1}$, $K'_4$=1.67 × $10^5$ $M^{-1}$.

TABLES

TABLE 1
Summary of Dependent and Independent Variables

| Variable | Equivalent | Comments |
|---|---|---|
| $B_1$ | $K_1CT_1$ | Fraction of Tm subunits coupled to the Tn-actin interaction and no calcium bound. Tm in Position B. |
| $B_2$ | $K_3CT_2$; $K_4B_1$ | Fraction of Tm subunits coupled to the Tn-actin interaction and one calcium bound. Tm in Position B. |
| $B_3$ | $K_5CT_3$; $K_4B_2$ | Fraction of Tm subunits coupled to the Tn-actin interaction and two calcium bound. Tm in Position B. |
| $B^-$ | $K_1CT^-$ | Fraction of Tm subunits coupled to the mutant Tn-actin complex. Tm in Position B. |
| $T_1$ | $1-B_1-B_2-B_3-T_2-T_3$; $p-B_1-B_2-B_3-T_2-T_3$ | Fraction of Tn dissociated from actin and tethered to Tm; no calcium bound. Tm position is undefined. |
| $T_2$ | $K_2T_1$ | Fraction of Tn dissociated from actin and tethered to Tm; one calcium bound. Tm position is undefined. |
| $T_3$ | $K_2T_2$ | Fraction of Tn dissociated from actin and tethered to Tm; two calcium bound. Tm position is undefined. |
| $T^-$ | $1-p-B^-$ | Fraction of mutant Tn dissociated from actin and tethered to Tm. |
| C | $1-M-B_1-B_2-B_3$ | Fraction of Tm directly associated with actin. Tm at equilibrium in Position C. |
| M | $K_0C(g-M)^n$ | Fraction of Tm coupled to the myosin-actin interaction. Tm in Position M. |
| $P_M$ | M | Probability of myosin coupled to the work associated with Tm in Position M; $0 \leq P_M \leq 1$ |
| Ca | | Calcium concentration; continuously independent variable. |

TABLE 2
Summary of Parameters

| Parameter | Value | Comments |
|---|---|---|
| $K_0$ | 1 | Composite of constants for myosin-actin interaction and coupling to the position of Tm; adjustable parameter for modeling. |
| $K_1$ | 500 | Composite of constants for Tn-actin interaction (no bound calcium) and coupling to the position of Tm; constrained by $K_1 > K_0$; adjustable parameter for modeling. |
| $K_3$ | $10K_1$ | Composite of constants for Tn-actin interaction (one bound calcium) and coupling to the position of Tm; value given by $K'_2/K'_4$ |
| $K_5$ | $100K_1$ | Composite of constants for Tn-actin interaction (two bound calcium) and coupling to the position of Tm; value given by $(K'_2/K'_4)^2$ |
| $K'_2$ | $1.67 \times 10^6 \, M^{-1}$ | Constant (evaluated from [14]) |
| $K'_4$ | $1.67 \times 10^5 \, M^{-1}$ | Constant (evaluated from [14]) |
| $K_2$ | $K'_2 Ca$ | Independent variable; allows input of calcium (Ca) for computation |
| $K_4$ | $K'_4 Ca$ | Independent variable; allows input of calcium (Ca) for computation |
| $\alpha$ | 3–6 | Maximum number of unsupported Tm subunits of a segment (evaluated from [22]) |
| $n$ | 1–5 | Number of coupled myosin per super segment; modeling results suggest $3 \leq n \leq 5$. |
| $p$ | 0–1 | Adjustable parameter of the fraction of native Tn |
| $g$ | $1 + \alpha P_M$ | Number of Tm subunits in a segment |
| $g_{max}$ | $1 + \alpha$ | Maximum segment length |
| $n(g_{max})$ | $n(1 + \alpha)$ | Maximum super segment length |

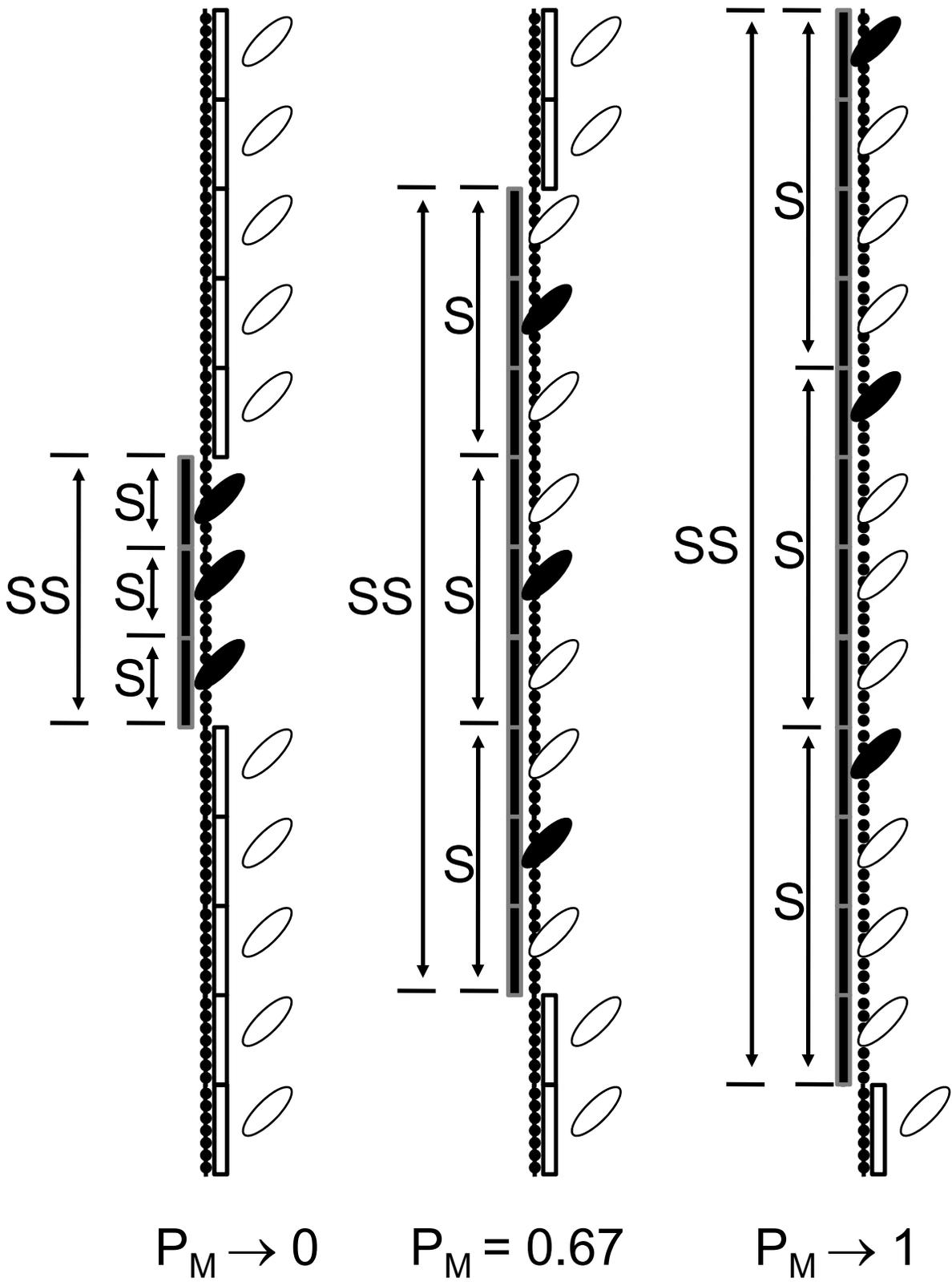

$P_M \rightarrow 0$     $P_M = 0.67$     $P_M \rightarrow 1$

Figure 1

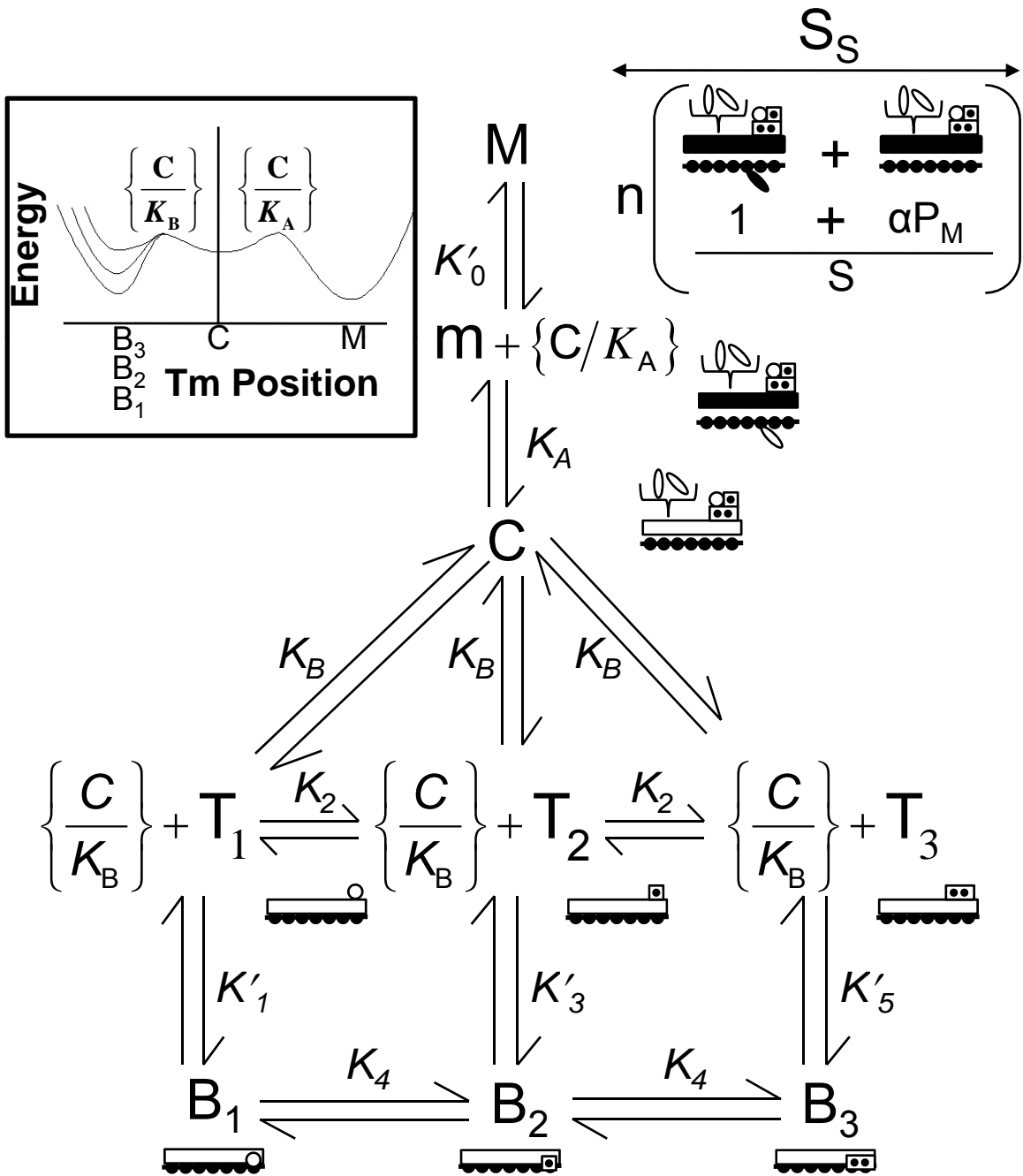

Figure 2

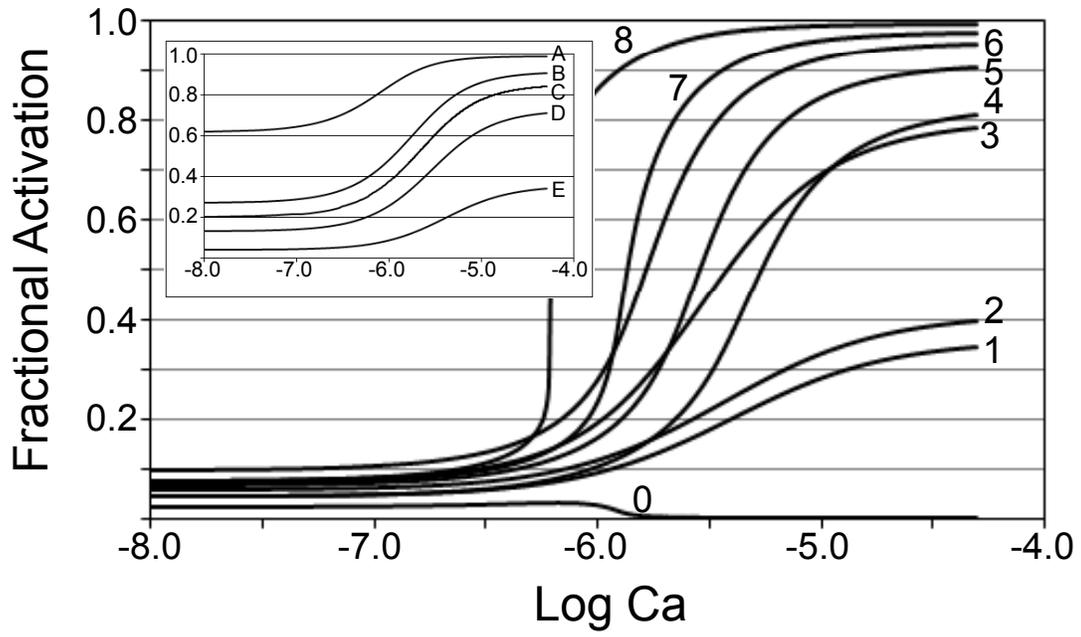

Figure 3

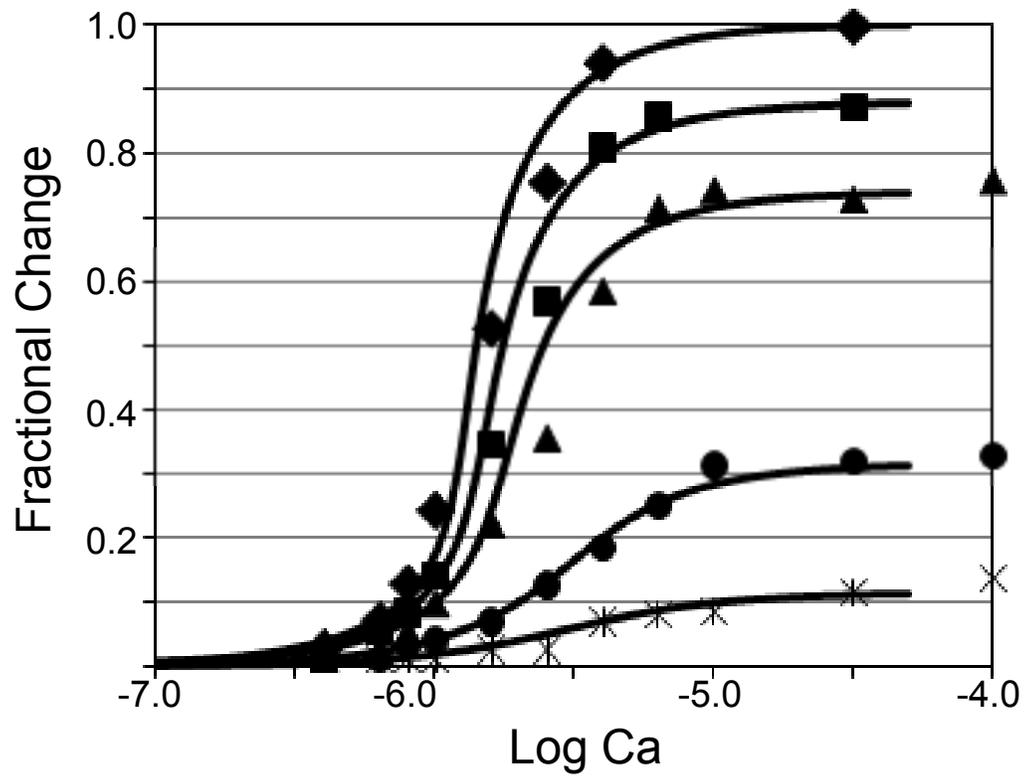

Figure 4

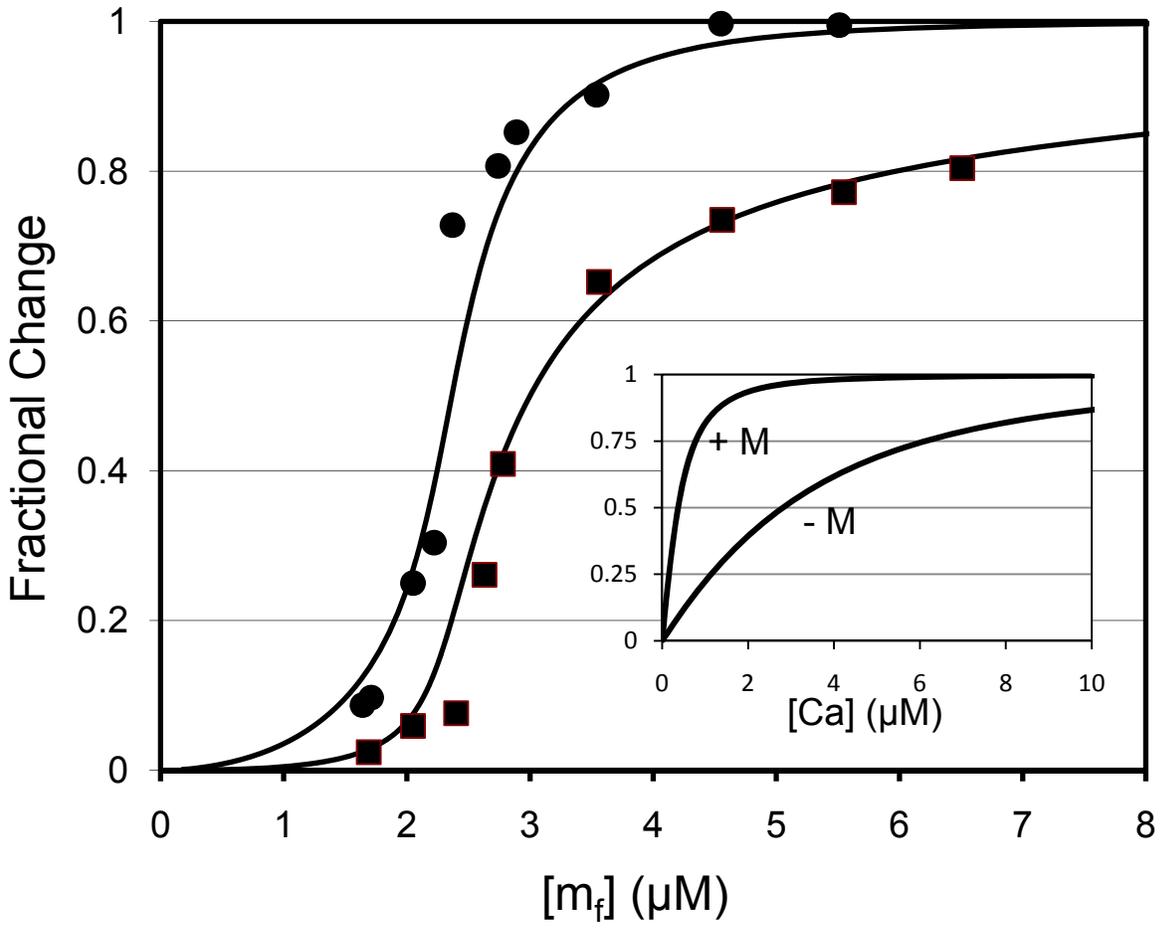

Figure 5